\documentclass[preprint,showpacs,preprintnumbers,amsmath,amssymb]{revtex4}

\usepackage{epsfig}
\usepackage{dcolumn}
\usepackage{bm}

\newcommand{\WIDTH}{4.5cm}  
\newcommand{\Width}{0.3cm}  
\newcommand{\qTEXT}[1]{\makebox[\WIDTH][c]{\small\textbf{#1}}}
\newcommand{\qtext}[1]{\makebox[\WIDTH][c]{\small #1}}
\newcommand{\qpic}[1]{\epsfig{file=q.#1.eps,width=\WIDTH,%
bbllx=0,bblly=0,bburx=360,bbury=220}}

\begin{document}

\preprint{HU--EP--01/54}
\preprint{RCNP--Th01030}
\preprint{ZIB--Report 01-33}
\preprint{revised manuscript}

\title{Parallel tempering in full QCD with Wilson fermions}

\author{E.-M.~Ilgenfritz}    
\affiliation{Research Center for Nuclear Physics, 
          Osaka University, Osaka 567-0047, Japan} 
\author{W.~Kerler}
\affiliation{Institut f\"ur Physik, Humboldt Universit\"at zu Berlin, 
          D-10115 Berlin, Germany}
\author{M.~M\"uller-Preussker}
\affiliation{Institut f\"ur Physik, Humboldt Universit\"at Berlin, 
          D-10115 Berlin, Germany}
\author{H.~St\"uben}
\affiliation{Konrad-Zuse-Zentrum f\"ur Informationstechnik Berlin, 
          D-14195 Berlin, Germany}

\date{13 February 2002}

\begin{abstract}
We study the performance of QCD simulations with dynamical Wilson
fermions by combining the Hybrid Monte Carlo algorithm with parallel
tempering on $10^4$ and $12^4$ lattices.  
In order to compare tempered with standard simulations, covariance matrices
between sub-ensembles have to be formulated and evaluated using the general
properties of autocorrelations of the parallel tempering algorithm.
We find that rendering the hopping parameter $\kappa$ dynamical does not 
lead to an essential improvement. We point out possible reasons for this 
observation and discuss more suitable ways of applying parallel 
tempering to QCD.
\end{abstract}

\maketitle

\section{\label{sec:intro}Introduction}

Improving Hybrid Monte Carlo (HMC) simulations of QCD with dynamical fermions 
is a long standing problem. While better decorrelation is, of course, highly 
desirable for all observables, it is of crucial importance for the ones which
are sensitive to topological sectors. In fact, it has been observed that the 
$\eta'$ correlator is definitely dependent on the topological charge $Q$
\cite{SESAM2}. Thus it is quite 
important to look for simulation methods that produce realistic 
$Q$-distributions. From this point of view the topological charge 
appears to be a good 
touchstone when looking for improvements by new methods. 

For staggered fermions an insufficient tunneling rate of the topological 
charge $Q$ has been observed \cite{MMP,Pisa}. For Wilson fermions the 
tunneling rate has been claimed to be adequate in many cases
\cite{SESAM,CP-PACS}. However, since the comparison is somewhat subtle,
the reason for this could also be that one is not as far in the critical 
region as with staggered fermions.
One could fear that simulating closer to the chiral limit, insufficient 
tunneling could become for Wilson fermions as severe as for staggered ones.
Indeed, for Wilson fermions on large 
lattices and for large values of $\kappa$ near the chiral limit the 
distribution of $Q$ is not symmetric even after more than 3000 trajectories 
(see {\it e.g.} Figure~1 of Ref.~\cite{SESAM}).

In the method of \emph{simulated} tempering first proposed in Ref.~\cite{mar92} 
the inverse temperature is made a dynamical variable in the simulations. 
More generally, any parameter in the action can be made dynamical.
Let us suppose that, depending on this particular coupling parameter, the chosen algorithm
has a largely different tunneling rate between certain metastable states (in configuration
space). Augmenting the algorithm with the tempering method means that now the system is 
updated in an enlarged configuration space including the coupling. Instead of overcoming 
a high barrier at an unfortunate parameter value, a detour in parameter space is now opened 
to be an easier route. This results in better decorrelation.

Considerable improvements 
have been obtained with dynamical number of the degrees of freedom in the 
Potts-Model \cite{ker93}, with dynamical inverse temperature for spin glass 
\cite{ker94} and with dynamical monopole coupling in U(1) lattice theory 
\cite{ker95}.  With dynamical mass of staggered fermions in full QCD \cite{boy97} 
it has been indicated that by tempering a better sampling of the configuration space 
(with respect to topological charge) can be achieved.
However, simulated tempering requires the determination of 
a weight function in the generalized action, and 
an efficient method of estimating it \cite{ker94,ker95} 
turns out crucial for successfully accelerating the simulation.

A major progress was the proposal of the \emph{parallel} tempering method (PT)
\cite{HN,EM}, in which no weight function needs to be determined. This method 
has allowed large improvements in the case of spin glass \cite{HN}. In QCD 
improvements have been reported with staggered fermions \cite{Boyd}, applying PT 
to subensembles characterized by different values 
of the quark mass. This has led to the expectation that, analogously 
in the case of Wilson fermions, introducing various values of the hopping 
parameter $\kappa$ might be the right choice of parameter for applying
the idea of PT. In a first study of this problem, 
simulations of QCD with O($a$)-improved Wilson fermions \cite{UKQCD}, 
no computational advantage has been found.
Because only two coupled sub-ensembles, both at relatively small $\kappa$,
had been used, this could not be the final answer concerning the potential
capabilities of the PT method. In a previous work \cite{us_1}, 
with more ensembles and (standard) Wilson fermions on an $8^4$ lattice, 
we have observed a considerable increase of the transitions between 
topological sectors. We have extended this study to larger lattices
($10^4$ and $12^4$) in \cite{us_2}.

In the present paper, our task will be to compare standard HMC with HMC 
combined with PT in a more elaborate, quantitative manner. 
In order to really compare algorithms
one has to relate the computational effort (computer time) to errors
of final results (e.g.\ particle masses).  In the case of PT
the calculation of errors becomes more complicated, 
because one has to take cross correlations
between ensembles into account.  Cross correlations lead to the
technical problem of calculating full covariance matrices from
(auto-) correlation functions. In this paper, we carry out such an analysis
for the average plaquette and the topological charge.

At the time when we were doing our autocorrelation analysis SESAM
published autocorrelation results for all their runs \cite{QME}.
Qualitatively the SESAM results are very similar to ours, i.e. they
also do not observe a clear increase of autocorrelation times with
increasing $\kappa$ (lower quark mass) in the interval
$[0.1560,0.1575]$.
One might have expected some mass dependence from the positive experience 
with tempering in the case of staggered fermions.
In the light of this fact, the final conclusion of our analysis, that the PT 
method with respect to the hopping parameter does not lead to an improved 
performance of the HMC algorithm, is not really astonishing. Nevertheless, 
formulating the tools for the correlation analysis is a major part of this 
paper which should prove useful for possible future applications of PT. 

The paper is organized as follows. In Section \ref{sec:method} 
we describe the method of PT to be applied to the HMC algorithm. 
Our simulation results are discussed in Section \ref{sec:results}. 
In Section \ref{sec:covariance} we give general properties of covariances and autocorrelations 
which -- in view of the moderate statistics available -- 
must be employed to fully exploit the autocorrelation data.
In Section \ref{sec:autocorr} we present and discuss our results on integrated autocorrelation times. 
The respective results for off-diagonal elements of the covariances are presented and 
discussed in Section \ref{sec:nondiagonal}. 
In Section \ref{sec:compare} we compare the efficiency of the two simulation 
algorithms (HMC without and with PT) and study the effect of cross correlations. 
In Section \ref{sec:accept} we comment on swap acceptance rates and give a new formula for that
rate extending the one used in Ref.~\cite{UKQCD}. We conclude in Section 
\ref{sec:conclusions} and point out potentially more promising applications of PT to QCD.

\section{\label{sec:method}Parallel tempering}

In standard Monte Carlo simulations one deals with one parameter set
$\alpha$ and generates a sequence of field configurations $F(s)$,
where $s$ denotes the Monte Carlo time. In our
case the set $\alpha$ includes the physical parameters $\beta$, $\kappa$ and
algorithmic parameters like the leapfrog time step and the number of
time steps per trajectory.  One field configuration $F(s)$ comprises the
gauge field and the pseudo fermion field.

In the parallel tempering (PT) approach \cite{HN,EM} one updates $K$ field 
configurations 
$F_{\nu}$ with $\nu = 1, \dots, K$ in the same run. The characteristic feature
is that the assignment of the parameter sets $\alpha_j$ with $j = 1, \dots, K$ 
to the field configurations $F_{\nu}$ changes in the course of a
tempered simulation. The total configuration at time $s$ thus 
consists of 
$B(s)$, $F_1(s)$, $F_2(s)$,..., $F_K(s)$ where the permutation 
\begin{equation}
B(s)=\left(\begin{array} 
{cccccc}\nu_1(s)&\nu_2(s)&\ldots&\nu_j(s)&\ldots&\nu_K(s)\\
           1    &    2   &\ldots&    j   &\ldots&    K \\
\end{array}\right)
\label{eq:permutation}
\end{equation}
describes the assignment of the field configurations $F_{\nu_j(s)}(s)$
to the parameter sets $\alpha_{j}$.
For short this approach is called PT with $K$ ensembles.

The update of the $F_{\nu}$ occurs by the usual HMC procedure using the 
parameter sets $\alpha_j$ as assigned at a given time. The update of $B$ is 
achieved by swapping pairs according to the Metropolis acceptance condition 
with probability
\begin{eqnarray}
P_{\rm swap}(i,j) &=& \min\left( 1, e^{-\Delta H} \right) \,,
\nonumber
\\
\Delta H &=& 
H(\alpha_i, F_{\nu_i}) + H(\alpha_j, F_{\nu_j})
- H(\alpha_i, F_{\nu_j}) - H(\alpha_j, F_{\nu_i}) 
\label{eq:Pswap-1}
\end{eqnarray}
where $H$ denotes the Hamiltonian of the HMC dynamics for the parameter set 
$\alpha_j$ and the field configurations $F_{\nu_j}$. The total update of the
Monte Carlo algorithm, after which its time $s$ increases by one, then 
consists of the updates of all $F_\mu$ followed by the full update of $B$ 
which consists of a sequence of attempts to swap pairs.

Detailed balance for 
the swapping follows from (\ref{eq:Pswap-1}). 
Ergodicity is obtained by updating 
all $F_{\nu}$ and by swapping pairs in such a way that all permutations 
(\ref{eq:permutation})
can be reached. There remains still the freedom of choosing the succession 
of the individual steps.   All such choices lead to legitimate algorithms, 
which might differ in efficiency.

Our choice of steps is such that the updates of all $F_{\nu}$ and that of $B$
alternate.  Our criterion for choosing the succession of swapping pairs in the 
update of $B$ has been to minimize the average time it takes for the 
association of a field configuration to the parameters to travel from the 
lowest to the largest $\kappa$-value. This
has led us to swap pairs belonging to neighboring $\kappa$-values and to
proceed with this from smaller to larger $\kappa$-values.

The observables of interest are associated to a specific parameter set 
$\alpha_j$.  We denote them as
\begin{equation}
{\cal O}_j(s) \equiv {\cal O}(F_{\nu_j(s)}(s)),\quad j=1,\ldots,K \,.
\label{eq:observables}
\end{equation}

\section{\label{sec:results}Simulation results}

We have simulated lattice QCD with standard
Wilson fermions and one-plaquette action for the gauge fields.
In the HMC program the standard conjugate gradient
inverter with even/odd preconditioning was used. The trajectory length was
always~1. The time steps were adjusted to get acceptance rates of about 70\%
in the HMC Metropolis step. In all cases 
(standard HMC, tempered runs for all lattice sizes and ensemble sizes)
1000 trajectories were generated, with additional
50--100 trajectories for thermalization.

We have performed tempered runs using 6 and 7 ensembles, all at $\beta = 5.6$ 
on $10^4$ and $12^4$ lattices, as well as standard HMC runs for comparison. 
Our simulations cover the $\kappa$-range investigated by 
SESAM ($\kappa = 0.156, 01565, 0.157, 0.1575$) \cite{SESAM}. 
In a large scale PT simulation analogous to that by SESAM one would be 
interested to use all subensembles (separated in $\kappa$) anyway, such 
that additional computational effort seems to be affordable. For instance, 
in the run with 6 ensembles we extended the 
$\kappa$-range by adding lower values of $\kappa$, while in the run with 7 
ensembles we have used a denser spacing of the 
$\kappa$-values. Our $\kappa$-values are listed in Table \ref{tab:autocorr}. 

The observables determined were the average plaquette $P$ and the 
topological charge $Q$. The topological charge was measured using its naively 
discretized plaquette form after doing 50 cooling steps of Cabibbo-Marinari 
type. 
This method gives multiples of a unit charge.  The value of the unit
charge is close to 1 and has to be determined from the measurements.

\begin{figure}[h]

\mbox{%
\qTEXT{Standard HMC}\hspace{\Width}%
\qTEXT{Tempered HMC}\hspace{\Width}%
\qTEXT{Tempered HMC}}

\mbox{%
\qtext{}\hspace{\Width}%
\qtext{6 ensembles}\hspace{\Width}%
\qtext{7 ensembles}}

\mbox{%
\qtext{}\hspace{\Width}%
\qtext{$\Delta\kappa = 0.0005$}\hspace{\Width}%
\qtext{$\Delta\kappa = 0.00025$}}

\bigskip

\mbox{\qpic{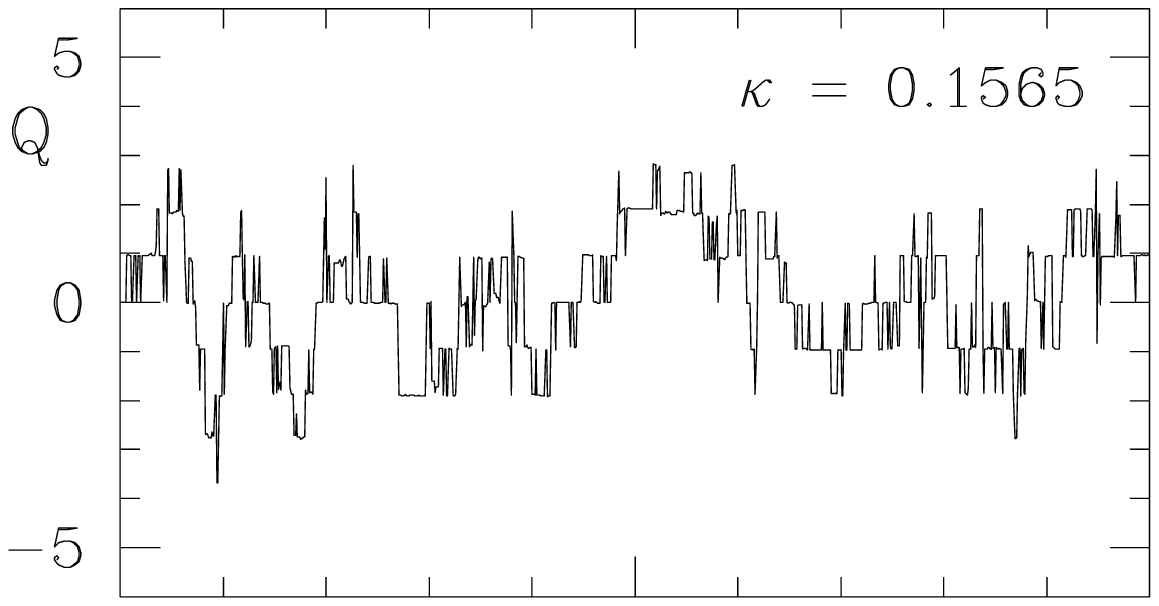}\hspace{\Width}\qpic{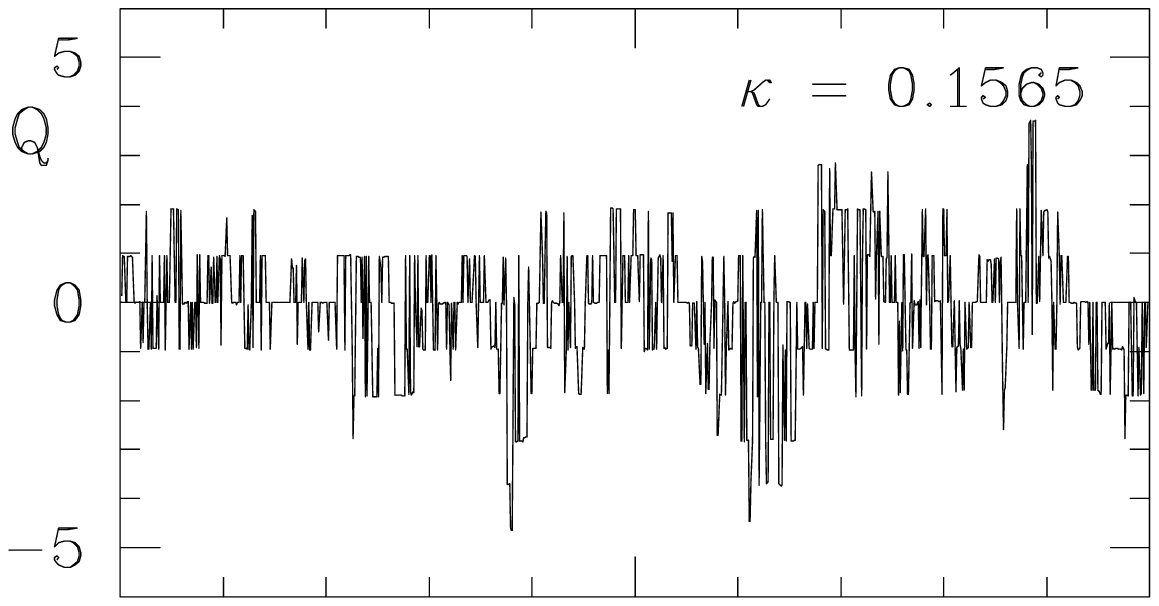}\hspace{\Width}\qpic{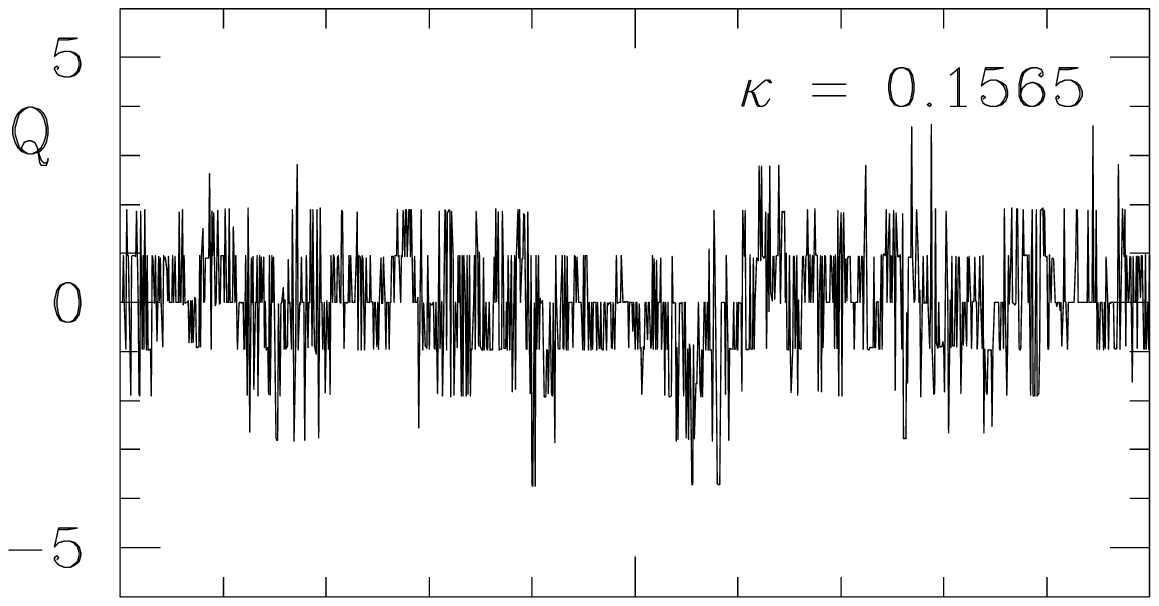}}
\mbox{\qpic{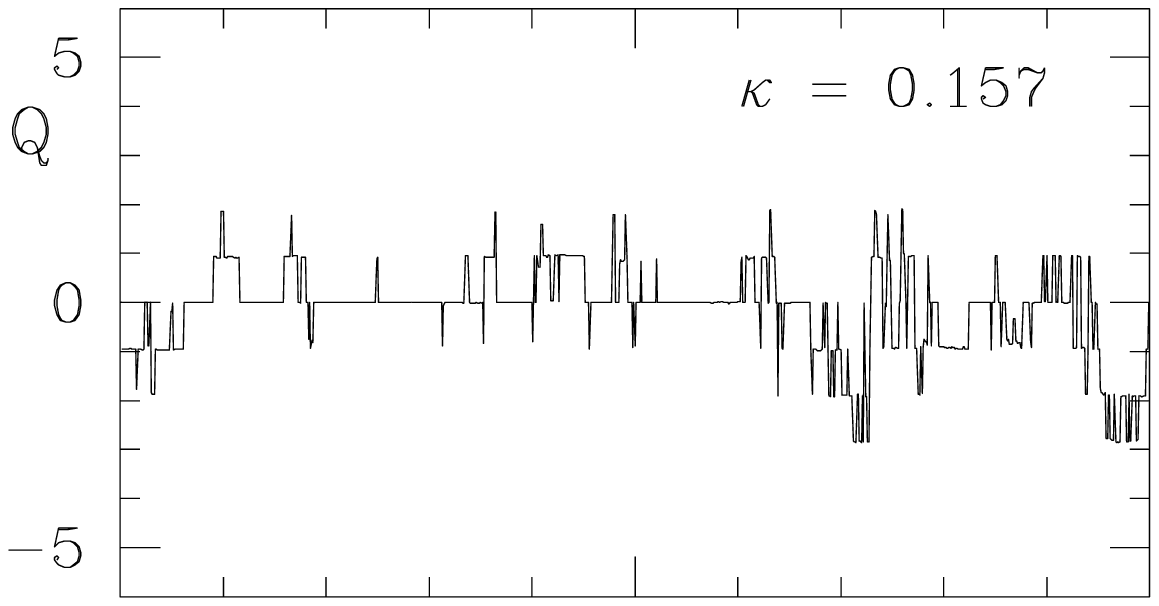}\hspace{\Width}\qpic{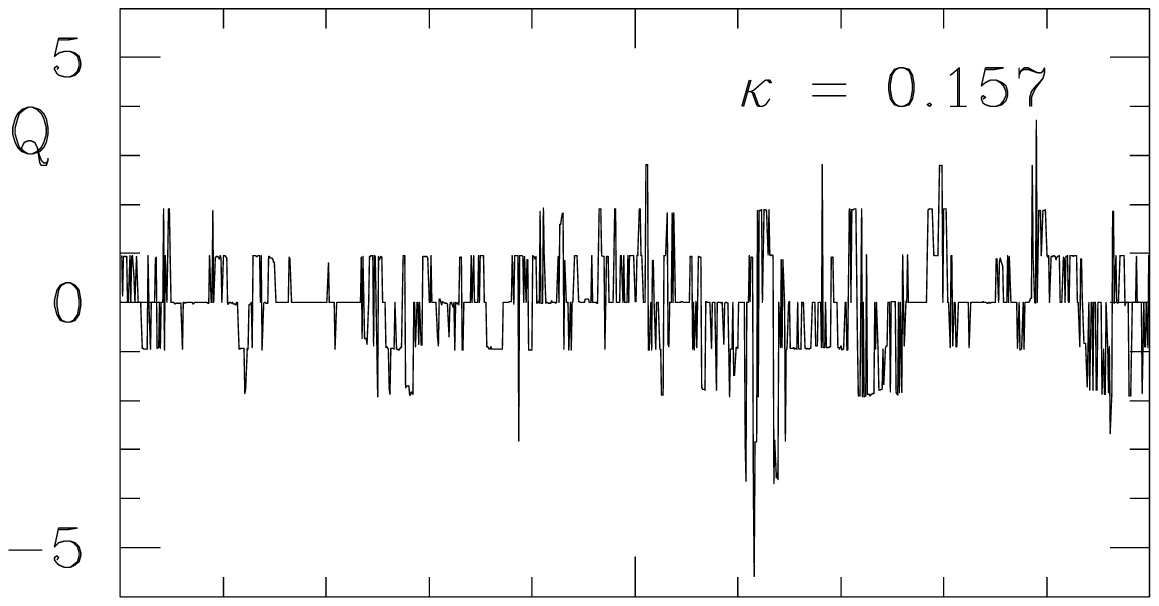}\hspace{\Width}\qpic{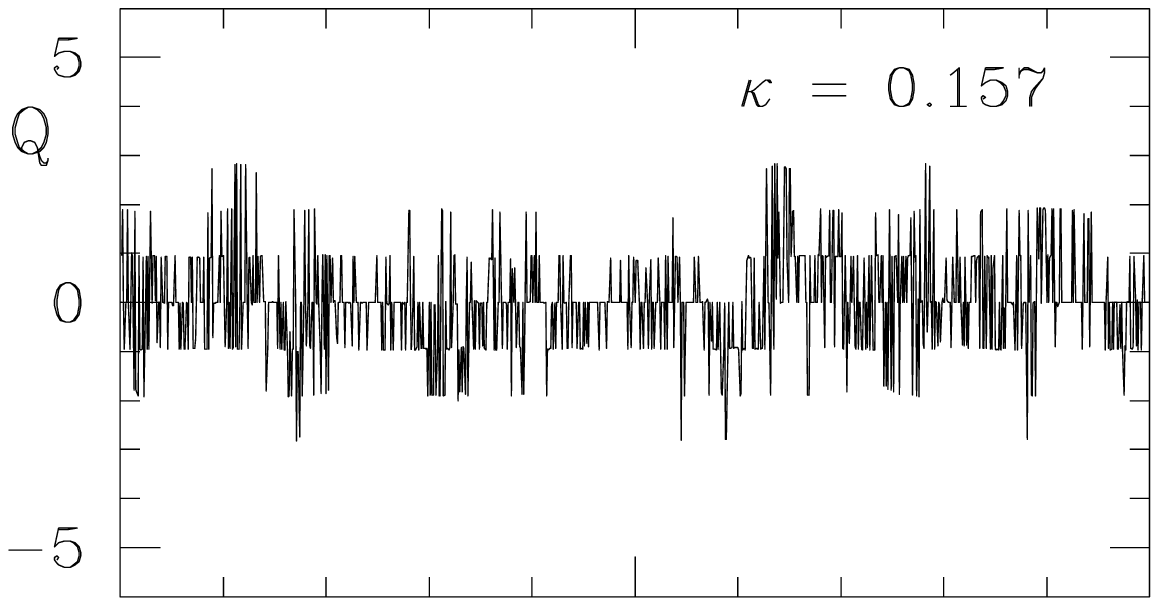}}
\mbox{\qpic{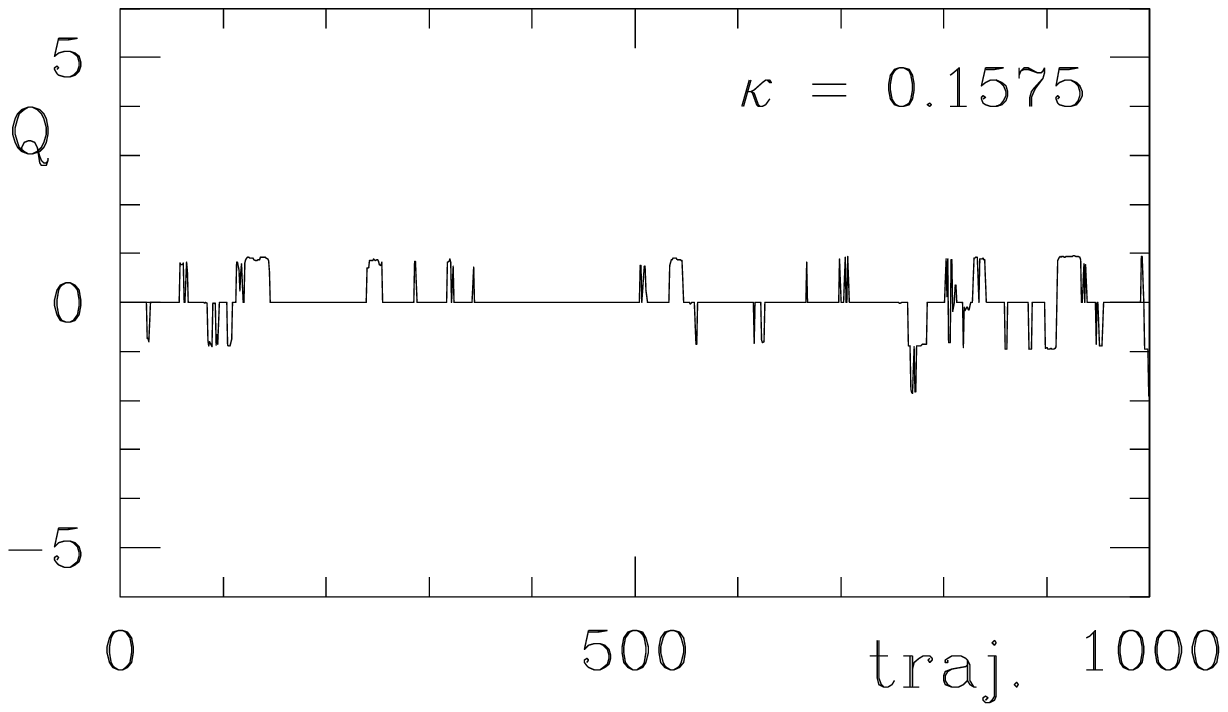}\hspace{\Width}\qpic{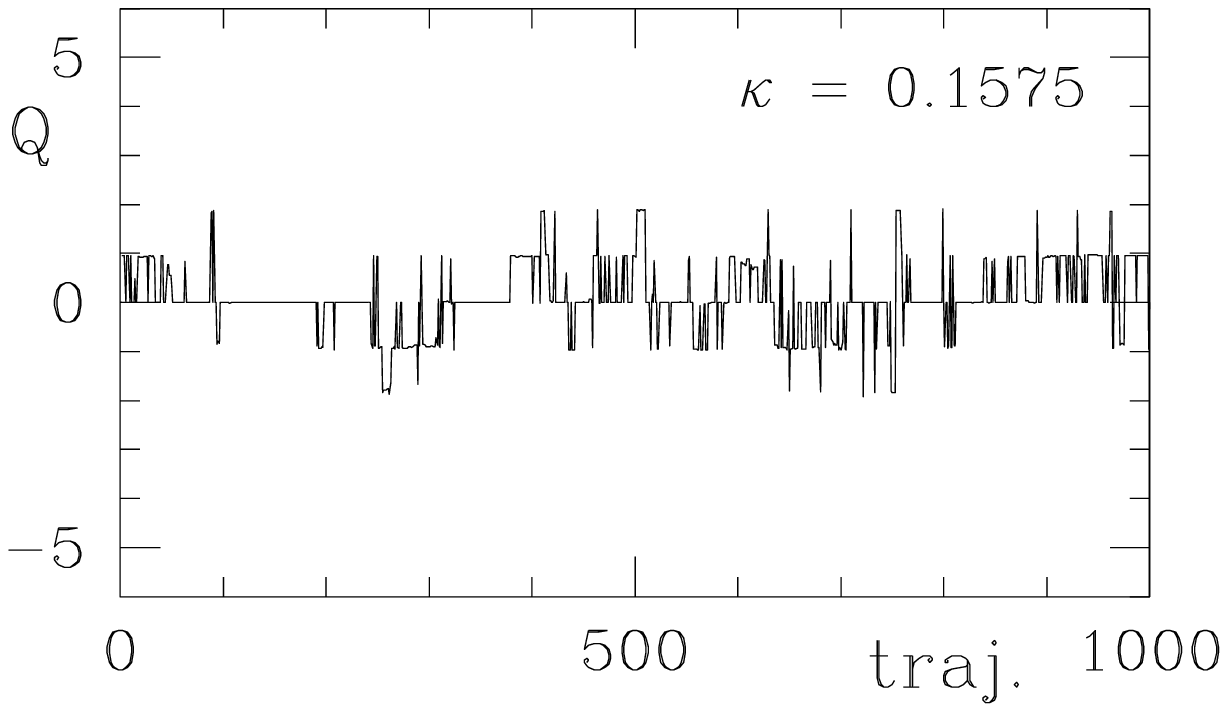}\hspace{\Width}\qpic{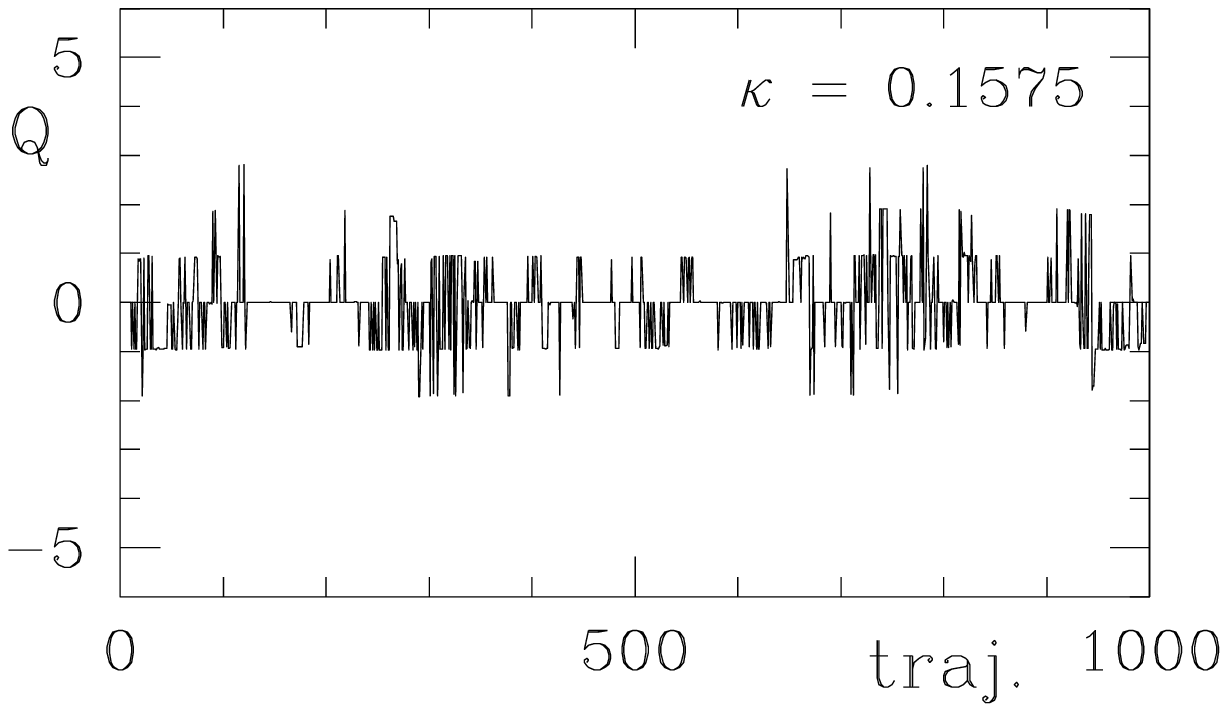}}

\bigskip

\caption{\label{fig:Q}Time series of $Q$ for standard and tempered HMC 
on  $12^4$ lattice at $\beta = 5.6$ (for part of 
$\kappa$-values only, see Table \ref{tab:autocorr} for full list).}
\end{figure}

Figure \ref{fig:Q} shows typical time series of $Q$ 
obtained for standard HMC
and for tempered HMC with 6 and with 7 ensembles.  One sees that
tempering makes $Q$ fluctuate much stronger. Such behavior is 
indicative for the 
decreasing of correlations between subsequent trajectories.
The time series on the $12^4$ lattice exhibits a richer pattern of
transitions as that on the $10^4$ lattice, and the width
of the topological-charge distribution increases. But the rate of 
fluctuations decreases with increasing $\kappa$.

In our previous investigations \cite{us_1,us_2} 
we have considered the mean absolute
change of $Q$ (called mobility in \cite{SESAM}) to account
quantitatively for these rates of fluctuations. However,
this quantity does not provide a quantitative measure of the computational
gain obtainable with the tempering method in comparison with standard 
HMC.

In the present investigation, we therefore base the comparison on 
the full account of the non-diagonal covariance matrix for different
observables to be introduced in Eq.~(\ref{eq:C-function}). 
The covariance matrix will be calculated from general correlation functions
which, for an observable ${\cal O}$ and a number $N$ of updates, are
defined as
\begin{equation}
R_{jk}(t)=\frac{1}{N}\sum_{s=1}^N {\cal O}_j(s) {\cal O}_k(s+t)-
\Big(\frac{1}{N}\sum_{s'=1}^N {\cal O}_j(s')\Big)
\Big(\frac{1}{N}\sum_{s''=1}^N {\cal O}_k(s'')\Big) \;.
\label{eq:R-function-1}
\end{equation}
For $j = k$ they are the usual autocorrelation functions, while for
$j \ne k$ they describe cross correlations between different ensembles.

Typical examples of normalized autocorrelation functions $\rho_j(t)=
R_{jj}(t)/R_{jj}(0)$ are presented in Figure \ref{fig:rho}. 
It can be seen that for the 
tempered runs the decay is considerably faster than for the 
standard ones. Among the tempering runs it is fastest for the run with 
7 ensembles.  In the latter case the remarkable fast decay 
occurring in the interval $t \in [0,1]$ should be noticed.

With the statistics available, the correlation functions decay 
below the Monte-Carlo noise for relatively small $t$. Although the much
faster decay of the functions for tempering is apparent, giving numbers
for the autocorrelation times and cross correlations is clearly a
formidable task. 
In principle, in order to estimate the autocorrelation times within
10 \% one would need trajectory numbers higher than ours by
roughly an order of magnitude. 
To reach a conclusion about the expected improvement we do not attempt 
such a precision measurement of the autocorrelation times.
Moreover, we want to apply some a priori knowledge on the spectral
properties in order
to exploit our simulation data in the most efficient way
possible. Therefore, in the following we first have to elaborate on some
theoretical issues concerning covariances.

\section{\label{sec:covariance}Covariance matrix and Markov spectrum}

We obtain the covariance matrix by using the general correlation
function (\ref{eq:R-function-1}) 
and generalizing the derivation given in Ref. \cite{pr89}
for the case $j=k$
\begin{equation}
C_{jk}=\frac{1}{N}\Bigg(R_{jk}(0)+\sum_{t=1}^{N-1}
\Big(1-\frac{t}{N}\Big)\Big(R_{jk}(t)+R_{kj}(t)\Big)\Bigg)\;.
\label{eq:C-function}
\end{equation}
The diagonal elements of (\ref{eq:C-function}) 
are the variances of ${\cal O}_j$ which are
traditionally written in the form 
\begin{equation}
\mbox{var}({\cal O}_j)=\frac{R_{jj}(0)}{N}\;2\tau_j \; ,
\label{eq:variance}
\end{equation}
introducing the integrated autocorrelation times 
\begin{equation}
\tau_j=\frac{1}{2}+\sum_{t=1}^{N-1}\rho_j(t) \; ,
\label{eq:tau-int}
\end{equation}
where $\rho_j(t)=R_{jj}(t)/R_{jj}(0)$. 

When evaluating
practical simulations the summation in (\ref{eq:tau-int}) 
up to $N-1$ makes no sense
since $\rho_j(t)$ is buried in the Monte Carlo noise already for relatively 
small $t$. Therefore, it has been proposed 
\cite{pr89,ma88} to sum up only to some smaller  value $M$ of $t$. 
However, in practice that procedure is not stable against the 
choice of $M$ and neglecting the rest is a bad approximation. The
proposal to estimate the neglect by an extrapolation based on the $t$ 
values $M$ and $M-1$ \cite{wo89} is still inaccurate in general. A more
satisfying procedure is to describe the rest by a fit function based on 
the (reliable) terms of (\ref{eq:C-function}) 
for $t \le M$ and on the general knowledge 
about the Markov spectrum. This procedure has led to perfect 
results in other applications \cite{ke93}.

In order to apply the latter strategy also for determining 
off-diagonal entries in (\ref{eq:C-function}) we have to look 
how spectral properties enter the parallel--tempering case. 
For such considerations it is convenient to introduce 
a Hilbert space \cite{ma88,ke93a}
with an inner product $(\phi,\chi)=\sum_{\cal C} 
\mu({\cal C}) \phi^*({\cal C}) \chi({\cal C})$, 
where $\mu({\cal C})$ is the equilibrium distribution of the system
and ${\cal C}$ denotes the configurations 
${\cal C} = \{ B,\;F_1,\;F_2,\ldots,F_K \}.$
Using this notation we can write the expectation values 
$\langle {\cal O}_j \rangle = \sum_{\cal C} \mu({\cal C}) 
{\cal O}_j({\cal C}) = \Big(1,{\cal O}_j\Big)$ 
and the two-time correlation functions 
\begin{equation}
\langle {\cal O}_j(0) {\cal O}_k(t) \rangle=
\sum_{{\cal C}^{[0]},{\cal C}^{[t]}}
    \mu({\cal C}^{[0]}) {\cal O}_j({\cal C}^{[0]})
W^t({\cal C}^{[0]};{\cal C}^{[t]}) {\cal O}_k({\cal C}^{[t]})=
\Big( {\cal O}_j, W^t {\cal O}_k \Big)
\label{eq:CC-corr}
\end{equation}
as inner products, where 
\begin{equation}
W^t({\cal C}^{[0]};{\cal C}^{[t]}) = 
\sum_{{\cal C}^{[1]},\ldots,{\cal C}^{[t-1]}}
W({\cal C}^{[0]} ;{\cal C}^{[1]}) W({\cal C}^{[1]};{\cal C}^{[2]})
\ldots W({\cal C}^{[t-1]};{\cal C}^{[t]}) 
\label{eq:multistep}
\end{equation}
is the $t$-step transition matrix constructed from
the one-step transition matrix $W({\cal C};{\cal C}')$ 
of the Markov process considered.
In the spectral representation 
\begin{equation}
W=\sum_{r\ge1}\lambda_r P_r
\label{eq:W-spectral}
\end{equation}
with eigenvalues $\lambda_r$ and projection 
operators $P_r$, one has $\lambda_1=1$ for the equilibrium eigenvector 
$\mu({\cal C})$ and 
$|\lambda_r|<1$ for the other modes. Obviously only $P_1$ contributes to the 
stationarity relation $\sum_{{\cal C}'} \mu({\cal C}') W({\cal C}';{\cal C})
=\mu({\cal C})$, and 
$P_1({\cal C}',{\cal C}) = \mu({\cal C})$ follows. With this notation
one can rewrite 
\begin{equation}
\langle {\cal O}_j \rangle \langle {\cal O}_k \rangle
= \Big( {\cal O}_j , P_1 {\cal O}_k \Big) \;.
\label{eq:uncorr}
\end{equation}
Using (\ref{eq:CC-corr}), (\ref{eq:W-spectral}) and (\ref{eq:uncorr}) 
we obtain for the general correlation 
function
\begin{equation}
R_{jk}(t)=\langle {\cal O}_j(0) {\cal O}_k(t) \rangle
-\langle {\cal O}_j \rangle
\langle {\cal O}_k \rangle 
=\sum_{r>1} \lambda_r^t \Big( {\cal O}_j, P_r {\cal O}_k \Big) \;,
\label{eq:R-function-2}
\end{equation}
where, due to the subtraction, the term with $\lambda_1=1$ cancels out. 
We thus get the general representation
\begin{equation}
R_{jk}(t)=\sum_{r>1}a_{jkr}\lambda_r^t \quad\mbox{with}\quad |\lambda_r|<1
\label{eq:R-function-3}
\end{equation}
where only 
the coefficients $a_{jkr}$ depend on the particular 
pair of observables while the eigenvalues $\lambda_r$ 
are universal and characterizing the chosen simulation algorithm.
It is important to realize that this also holds for the observables of form 
(\ref{eq:observables}) used in PT.

\section{\label{sec:autocorr}Autocorrelation results}

For the numerical evaluation of (\ref{eq:C-function}) 
we apply the method explained
in Section \ref{sec:covariance} using the fact that after some time only the
slowest mode in (\ref{eq:R-function-3}) survives.  Our method is to sum up the
simulation data only up to some $t$ before the noisy region and
determining the rest of the sum from a fit assuming that the fit
describes the slowest mode well.  For $\tau_j$ the rest typically
amounts maximally to about 25 \%.  The proper choice of the fit
intervals in $t$ (excluding the region of fast contributions and the
noisy region) was controlled by inspection of the graphs and watching
the resulting $\chi^2$ values.  Examples of such fits are shown in
Figure \ref{fig:rho}.

In view of the moderate statistics available we additionally have made
use of the universality of the Markov spectrum implying that in
(\ref{eq:R-function-3}) for given algorithm and lattice size only the coefficients
$a_{jkr}$ can vary with the observables.  We have verified that a
collective fit for the whole diagonal with a universal slowest mode
gives results comparable to individual fits (see Table \ref{tab:autocorr}).  
That motivated us to perform collective fits with one single mode to
diagonal and non-diagonal terms.  In fact that method greatly helped
to get stable fits which will be further discussed in section
\ref{sec:nondiagonal}.

To obtain errors 
for the covariances one can generalize the derivations given in 
Ref. \cite{pr89} for the diagonal case to calculate covariances of covariances 
from the $R_{jk}(t)$ data only.  However, such calculation is
impractical with the statistics available. Therefore, we have to rely on 
the comparison of measurements of covariances at different parameter values 
and on consistency checks to get some idea about the size of the errors. 

Comparing the results for $R_{jj}(0)$ from the different simulation algorithms,
which should give the same numbers, one sees that this ingredient of
a calculation of $\tau_j$ has errors of about 20 \%. The exponential 
autocorrelation time $\tau_{\rm exp}=-1/{\rm ln }\lambda$, where $\lambda$ 
denotes the slowest mode, corresponds within a good approximation to the 
integrated autocorrelation time which one would get taking only the slowest 
mode into account.  The presence of faster modes then renders $\tau_j$ smaller
than $\tau_{\rm exp}\,$.  We always find consistency with this requirement. 

The integrated autocorrelation times $\tau_j$ obtained in this way are
given in Table \ref{tab:autocorr}.  One can see that there is little
difference between the results from individual and collective fits.
The fluctuations of values found for neighboring $\kappa$-values indicate
relatively large errors. Judging from the observed noise levels
the errors are expected to be largest for the standard case and smallest for 
tempering with 7 ensembles. Despite these errors two unexpected 
features are clearly visible: i) there is no sizable increase 
in $\tau_j$ with $\kappa$ and ii) there is nevertheless gain in 
terms of $\tau_j$ when using tempering.

The lack of a sizable increase in $\tau_j$ with $\kappa$, which one would
have expected for the standard runs, firstly indicates that valuing time 
histories by eye (as of $Q$ in Figure \ref{fig:Q}) 
can be misleading. It secondly
shows the important fact that the usual precondition of 
successful tempering, connecting regions with considerably different $\tau$, 
is not fulfilled here.

In the light of this it comes with some surprise that nevertheless 
gain in terms of $\tau_j$ is observed. 
The reduction of $\tau_j$ for $Q$ turns out to be larger than for $P$.

\section{\label{sec:nondiagonal}Off-diagonal covariances}

For the discussion of cross correlations in the following Section we need
the off-diagonal elements of the covariance matrices. As in the
diagonal case, the use of the simulation data in the sum (\ref{eq:C-function}) 
makes
only sense outside the noisy region.  One finds that the off-diagonal
elements of the general correlation functions are decreasing with the
distance from the diagonal.  Therefore, the evaluation of the sum
(\ref{eq:C-function}) 
in the off-diagonal case is more difficult because beyond some
distance $|j-k|$ the elements are completely indiscernible within the
noise.  See Figure \ref{fig:R} for an illustration.

In tempering with 7 ensembles we generally find that three off-diagonals
can be determined while for tempering with 6 ensembles only one. For the
off-diagonal elements $R_{jk}(t)$ which clearly show a signal above
the noise we generally observe a maximum at $t=|j-k|$ (see Figure \ref{fig:R}).
We therefore look for a prediction of their functional form. Rewriting
(\ref{eq:R-function-2}) 
as $R_{jk}(t)=\Big( {\cal O}_j(s), (1-P_1) W^t {\cal O}_k(s) \Big)$ their
qualitative behavior can be discussed. Obviously at each time the 
observables of type (\ref{eq:observables}) 
depend only on one field configuration.
This form of $R_{jk}(t)$ suggests that for $j\ne k$ a sizable
contribution only arises when, under the action of the transition
matrix by $W^t {\cal O}_k(s)$, a contribution also depending on the field 
configuration entering ${\cal O}_j(s)$ has been generated.  For our type of
swapping this situation occurs for $t \ge |j-k|$ so that 
\begin{equation} 
R_{jk}(t) \approx \left \{\begin{array} {cl}\sum_{r>1}\tilde{a}_{jkr}\lambda_r^t
&\mbox{for}\quad|j-k|\le t\\0&\mbox{for}\quad0\le t<|j-k| \end{array}\right\}
\quad \mbox{for}\quad j\ne k
\label{eq:Wtc}
\end{equation}
should be the approximate behavior. We indeed generally see this behavior
within errors in our data.

For the numerical evaluation of (\ref{eq:C-function}) we again apply the method
explained in Section \ref{sec:covariance} restricted to a $t$ interval where the
respective $R_{jk}(t)$ signal is sufficiently above the noise.  To get
stable off-diagonal results we use the method of the collective fit,
using the existence of a universal slowest mode described in Section
\ref{sec:autocorr}. Table \ref{tab:covmat} shows an example of a numerical 
result for
$C_{jk}$ (remember that $C_{jk}$ is symmetric). Generally the
off-diagonal elements obtained, especially the smaller ones, are
likely to be overestimated because of possible contributions of the
noise.

In the applications to be described in Section \ref{sec:compare} the full
covariance matrix $C_{jk}$ is needed. This excludes, for this purpose,
the consideration of the PT
results in our case of 6 ensembles, because we were able to determine elements 
only in the first
subdiagonal $|j-k|=1$.  They have much larger errors than in our PT 
studies with 7 ensembles.  In this case, in
contrast, we got 3 sub-diagonals.  We find that putting
the remaining ones equal to zero or using various extrapolations in
$|j-k|$ for them makes only little difference in the results. This
reflects the fact that, although the elements close to the diagonal
are not small, there is nevertheless a faster decrease farther away
from the diagonal.

\section{\label{sec:compare}Comparison of simulation algorithms}

At the end of Section \ref{sec:autocorr} we have pointed out that there is no increase
of autocorrelation times with $\kappa$. Because of this observation the
usual mechanism of tempering -- which is to  provide an easier detour 
through parameter space for the suppressed transitions -- is not available
here. If this mechanism is working, tempering with several parameter points
can be advantageous even if one is interested only in the result at one point.
This holds, in particular, for systems where in the region of interest
otherwise almost no transition occurs \cite{ker93,ker95}. Unfortunately 
our region is not of this type. 

As also observed in Section \ref{sec:autocorr} there is nevertheless a reduction of 
autocorrelation times. The effect of
this is, however, not large enough to get generally gain if one is interested
in only one point. In fact, dividing the reduction factors of the $\tau_j$
in Table \ref{tab:autocorr} 
by the number of ensembles it can be seen that at best in the case
of $Q$ some gain remains.

We now turn to the question whether gain remains if one is interested in the 
results at all parameter values. In this case it is necessary to account for
cross correlations between the ensembles. To be able to do this one has to
rely on fits to the data. The respective fit method is well know from the 
treatment of indirect measurements (see e.g.~Ref.~\cite{br92}). For proper
comparison this method, which leads to improved errors, has to be applied
to the tempering case as well as to the standard case.

Final results then are obtained from fits to mean values from individual 
ensembles.  In the case of PT the full covariance matrix 
enters the fit.  Although there are difficulties to account for the 
full matrix numerically, we have tried to develop some feeling for its 
influence by making fits to the observables we have measured.

It is known that $\langle Q \rangle = 0$. Therefore our fit ansatz for
$\langle Q \rangle$ is a constant, i.e., the fit procedure is a
weighted mean using the full covariance matrix.  For the plaquette we
observed that our data are consistent with a linear dependence on
$\kappa$.  Therefore we have used a linear fit ansatz in $\kappa$.

In the following we outline the fit method for the case of the plaquette.  
The linear fit ansatz just mentioned is $\langle P
\rangle = x_1 + x_2 \kappa$, where $x_1$ and $x_2$ are the fit
parameters.  In matrix notation we have $\eta = -Ax$ with
\begin{equation}
\eta = \left( \begin{array}{c}
\langle P \rangle_1 \\ \vdots \\ \langle P \rangle_K \\
\end{array} \right), \quad
A = - \left( \begin{array}{cc}
     1 & \kappa_1 \\
\vdots & \vdots   \\
     1 & \kappa_K \\
\end{array} \right), \quad
x = \left( \begin{array}{c} x_1 \\ x_2 \\ \end{array} \right).
\end{equation}
We also introduce the vector of measured values $p = (\bar{P}_1,
\dots, \bar{P}_K)^{T}$ and call the corresponding covariance matrix
$C_p$. The result of the fit is the minimum of $(\eta - p)^{T} C_p^{-1}
(\eta - p)$ which lies at $\tilde{x} = -(A^{T} C_p^{-1}A)^{-1}A^{T}
C_p^{-1}p$.  The errors of the result are square roots of the diagonal
entries of $C_{\tilde{x}}= (A^{T} C_p^{-1}A)^{-1}$.  For fitting a
constant to measured values of $\langle Q \rangle$ corresponding
formulae apply with $x_2 \equiv 0$.

Actually we are interested in the variances of the fit-function values. 
To get them we insert $\tilde{x}$ into the fit function, 
\begin{equation}
\tilde{\eta}=-A\tilde{x} \;,
\label{eq:eta}
\end{equation}  
and using (\ref{eq:eta}) in the transformation law of covariances we obtain
\begin{equation}
C_{\tilde{\eta}}=AC_{\tilde{x}}A^{T} \;.
\label{eq:C-eta}
\end{equation}
The diagonal elements of (\ref{eq:C-eta}) are the variances of interest and
the square roots of them the improved errors. 

For standard HMC, in these calculations we have used
the measurements for the 5 selected $\kappa$ values 
(see Table \ref{tab:autocorr}).  
In the case of tempering the
number of measurements is equal to the number $K$ of ensembles. 

Table \ref{tab:error} 
gives data with usual statistical errors and fit results (\ref{eq:eta}) 
with  improved errors (\ref{eq:C-eta}). 
We denote the errors by $e_0$ for the usual statistical errors (in column 2),
by $e_f$ for the improved errors taking into account the {\it full} covariance 
matrix (column 3), and by $e_d$ for those obtained only with the {\it diagonal} 
elements of the covariance matrix (column 4), respectively.   

The factor $(e_f/e_d)^2$ describes the influence of cross correlations.
We typically find values of about 2 to 3 for it. As compared to the
reduction factors apparent from Table \ref{tab:autocorr} 
this appears not large. However,
one has to be aware that proper comparison here needs consideration of
the improved errors in the standard {\it and} in the tempering case.

The relevant computational gain
factor for the comparison standard {\it vs.}~tempering case 
is given by the improved errors and the numbers of ensembles as
\begin{equation}
(e_d^{standard}/e_f^{tempering})^2 \; N_{standard}/N_{tempering} \;.
\label{eq:errors}
\end{equation}
In the example in Table \ref{tab:error} 
this factor appears close to one. However,
because of the inaccuracy of the standard data the respective fit results
are not reliable (giving factors from about 0.5 to 4 in other cases).
Thus we are not able to give definite numerical results for (\ref{eq:errors}).

A further quantity to be mentioned is the reduction factor $e_0/e_f$ for
the errors. This reduction in general is larger for larger $e_0$. Thus
the comparison based on the fit procedure favors the standard case, which
reduces a possible gain. 

Altogether it looks that even with more accurate data it might be difficult
to present evidence for computational gain in our case, where the 
autocorrelation times for standard HMC do not vary within the
$\kappa$-range considered.

\section{\label{sec:accept}Swap acceptance rates}

For the effect of enlarging the lattice size 
on the efficiency of the algorithm the change of 
the swap acceptance rate 
$\langle P_{\rm swap}\rangle$ resulting from (\ref{eq:Pswap-1}) appears to be relevant. 
In Ref.~\cite{UKQCD} agreement has been reported between the behavior observed in 
PT with the expression $\mbox{erfc} 
\left( \frac{1}{2} \sqrt{\langle \Delta H \rangle} \right)$ derived in 
Ref.~\cite{Bielefeld}.
The derivation given there,  
however, relies on the area-preserving property of the HMC 
algorithm, implying $\langle \exp(-\Delta H)\rangle=1$, which does not hold
in the case of swapping. 

To get the relation appropriate for swapping, we again neglect higher cumulants
in the cumulant expansion
$\langle \exp(-\Delta H)\rangle=\exp\Big(-\langle \Delta H\rangle +\frac{1}{2}
\langle(\Delta H-\langle \Delta H\rangle)^2\rangle)\mp\ldots\Big)$.
However, in contrast to Ref.~\cite{Bielefeld}, we
put $\langle \exp(-\Delta H)\rangle=\exp(-\delta)$ with an unknown $\delta$.
By convexity of the exponential function one finds that 
$\langle \Delta H\rangle \ge \delta$ holds. The relation between mean and 
width of the gaussian used in Ref.~\cite{Bielefeld} then generalizes to
\begin{equation}
\langle\Delta H\rangle=\frac{1}{2}\sigma^2+\delta \quad,\quad 
\sigma^2=\langle(\Delta H-\langle \Delta H\rangle)^2\rangle \; .
\end{equation}
The evaluation of the integral 
$ \frac{1}{\sqrt{2\pi}\sigma}\int_{-\infty}^{\infty}{\rm d}x\;\min(1,\exp(-x))
\exp\Big(-\frac{(x-\langle\Delta H\rangle)^2}{2\sigma^2}\Big)$ gives then 
\begin{equation}
\langle P_{\rm swap}\rangle=
\frac{1}{2}{\rm erfc}\Bigg(\frac{1}{2}\Big(u+\frac{\delta}{u}\Big)\Bigg)
+\frac{\exp(-\delta)}{2}{\rm erfc}\Bigg(\frac{1}{2}\Big(u-\frac{\delta}{u}
\Big)\Bigg)
\label{eq:Pswap-2}
\end{equation}
where $u=\sqrt{\langle \Delta H\rangle-\delta }$. 

Within errors our values of $\langle\Delta H\rangle$ turn out to scale with 
the volume $L^4$ (being roughly 1.4 and 2.8 for 6 ensembles and 0.35 and 0.7 
for 7 ensembles for $10^4$ and $12^4$, respectively). While the values of 
$\langle\exp(-\Delta H) \rangle$ for the $10^4$ lattice within errors conform 
with $1$, on the $12^4$ lattice they deviate substantially from this 
(increasing from 0.6 to 1.4).

Our data agree with (\ref{eq:Pswap-2}) using $\delta$ as found from our simulations. In 
the cases where we find that $\delta=0$ is not true, despite of this using 
$\delta=0$ in the acceptance formulae within errors gives still consistency. 
However, for larger lattices, where further increase of the deviations of 
$\langle\exp(-\Delta H) \rangle$ from 1 is to be expected, this might be no
longer so. The indicated deviations tend to improve the situation on larger 
lattices. Since the behavior of $\delta$ is not known quantitatively, 
detailed predictions on $\langle\Delta H\rangle$ at present are not possible.

\section{\label{sec:conclusions}Conclusions} 

In this paper we have compared PT with
standard HMC quantitatively on $10^4$ and $12^4$ lattices.  We have
described the steps of such an analysis and carried them out for the
average plaquette and the topological charge.  In a quantitative
analysis one has to look at the size of errors and therefore in
principle also at errors of the errors. 
While the first is at the limits of feasibility, 
the latter is definitely beyond our statistics.  
For that purpose we have demonstrated how the cross-correlation functions
between the sub-ensembles have to be taken into account, using their
general, algorithm independent properties. This has allowed us 
to make consistency checks, and we found consistent behavior of the results.
We believe that this part of our analysis is of general interest.

The choice of the $\kappa$-range in this paper was
guided by recent large-scale QCD simulations with
dynamical Wilson fermions like SESAM \cite{SESAM} and
triggered by the suppression of tunneling rates for the topological charge 
observed at the largest $\kappa$-values available. However, this range  
might be still too far from the chiral limit with
the consequence of no dramatic change of the autocorrelation times
for the standard Hybrid Monte Carlo method.  This would explain why 
PT in our case, 
{\it i.e.} for the $\kappa$ range covered by the work of SESAM,
did not provide a considerable computational gain.

At stronger coupling we know that approaching the chiral limit we
arrive at a second order transition into a phase with broken combined 
parity-flavor symmetry (so-called Aoki-phase) \cite{aoki}. At this transition
we expect a critical slowing down and therefore strongly increasing autocorrelations.
This is the reason, why we nevertheless believe that PT should become efficient 
if one is sufficiently close to the chiral limit.  
An even more promising scenario might be the application of the PT method towards the
continuum limit, in particular along lines of constant physics in the $\beta-\kappa$-plane. 
Already in pure $SU(3)$-Yang-Mills theory the tunneling rate between different topological 
sectors becomes strongly suppressed with increasing $\beta$.

The tools developed in the present paper will be very useful for such
future applications.

\begin{acknowledgments}
The simulations were done on the CRAY T3E at
Konrad-Zuse-Zentrum f\"ur Informationstechnik Berlin.
E.-M. I. gratefully appreciates the support by the Ministry
of Education, Culture and Science of Japan (Monbu-Kagaku-sho) 
and thanks H. Toki for the hospitality at RCNP.
\end{acknowledgments}



\newlength{\lengthPoint}
\newlength{\lengthDigit}
\settowidth{\lengthPoint}{.}
\settowidth{\lengthDigit}{0}
\newcommand{\sP}{\makebox[\lengthPoint]{}}  
\newcommand{\sD}{\makebox[\lengthDigit]{}}  

\begin{table*}[h]
\caption{\label{tab:autocorr}Integrated 
autocorrelation times $\tau_{int,P}$ for plaquette
and $\tau_{int,Q}$ for topological charge from individual fits (see 
Section \ref{sec:autocorr}). For comparison results from 
collective fits are given in brackets. 
For each case 1000 trajectories were generated with trajectory 
length 1.}
\vspace*{4mm}

\newlength{\X}
\settowidth{\X}{$-$}
\newcommand{\M}{\makebox[\X]{$-$}}  
\renewcommand{\P}{\makebox[\X]{}}   

\begin{tabular*}{16cm}{@{\extracolsep{\fill}}rrrrrrr}

\hline
&
\multicolumn{2}{c}{standard HMC} &
\multicolumn{4}{c}{tempered HMC} \\

& & &
\multicolumn{2}{c}{6 ensembles} &
\multicolumn{2}{c}{7 ensembles} \\

& & &
\multicolumn{2}{c}{$\Delta\kappa = 0.0005$} &
\multicolumn{2}{c}{$\Delta\kappa = 0.00025$} \\

\cline{2-3}\cline{4-5}\cline{6-7}
&
\multicolumn{1}{c}{$10^4$} &
\multicolumn{1}{c}{$12^4$} &
\multicolumn{1}{c}{$10^4$} &
\multicolumn{1}{c}{$12^4$} &
\multicolumn{1}{c}{$10^4$} &
\multicolumn{1}{c}{$12^4$} 
\\ \hline \\

\multicolumn{1}{c}{$\kappa$} & \multicolumn{6}{c}{$2\tau_{{\rm int},P}$} \\
\hline

0.15500 & 14.0  & 16.7  & 5.9 [6.7] & 8.8 [9.5] &       &       \\
0.15550 &       &       & 6.3 [6.9] &10.2 [8.6] &       &       \\
0.15600 & 10.4  & 13.3  & 3.7 [4.0] & 5.5 [6.2] & 4.2 [4.6] & 4.6 [5.1]\\
0.15625 &       &       &           &           & 2.8 [2.6] & 2.8 [3.4]\\
0.15650 & 9.3   & 12.9  & 5.2 [5.6] & 6.1 [6.9] & 2.9 [2.5] & 3.4 [3.4]\\
0.15675 &       &       &           &           & 2.7 [2.4] & 4.5 [3.7]\\
0.15700 & 8.6   & 14.2  & 5.8 [5.6] & 7.8 [8.1] & 2.9 [2.4] & 6.1 [4.8]\\
0.15725 &       &       &           &           & 4.1 [2.9] & 5.2 [4.9]\\
0.15750 & 9.0   & 8.1   & 8.4 [8.5] &10.9 [10.8]& 3.4 [3.9] & 8.2 [7.9]\\
\hline \\ $~$ \\
\multicolumn{1}{c}{$\kappa$} & \multicolumn{6}{c}{$2\tau_{{\rm int},Q}$} \\
\hline

0.15500 & 42    & 22    & 15   [17] & 17 [23] &      &       \\
0.15550 &       &       & 13   [11] & 11 [12] &      &       \\
0.15600 & 37    & 74    &  7 \sD[7] & 20 [19] & 16.6   [16.5] & 4.8 [7.7] \\
0.15625 &       &       &           &         &  9.7 \sD[9.7] & 5.2 [6.3] \\
0.15650 & 41    & 48    &  8 \sD[7] & 16 [17] &  5.7 \sD[6.1] & 5.8 [6.5] \\
0.15675 &       &       &           &         &  3.3 \sD[3.1] & 6.4 [6.8] \\
0.15700 & 45    & 38    & 16   [12] &  8 [10] &  2.3 \sD[2.5] & 6.3 [6.5] \\
0.15725 &       &       &           &         &  5.3 \sD[4.9] & 5.3 [5.5] \\
0.15750 & 46    & 14    &  6 \sD[8] & 35 [26] &  6.8 \sD[7.6] & 5.5 [7.0] \\
\hline

\end{tabular*}

\end{table*}

\vspace*{8mm}


\begin{table*}[h]
\caption{\label{tab:covmat}Covariance matrix of $Q$ for 
tempering with 7 ensembles on $12^4$ 
lattice.}

\vspace*{4mm}

\begin{tabular*}{16cm}{@{\extracolsep{\fill}}rrrrrr}

\hline
\multicolumn{1}{c}{$j$} & 
\multicolumn{1}{c}{$\kappa_j$} & 
\multicolumn{1}{c}{$10^3\cdot R_{j,j}$} & 
\multicolumn{1}{c}{$10^3\cdot R_{j,j+1}$} &
\multicolumn{1}{c}{$10^3\cdot R_{j,j+2}$} &
\multicolumn{1}{c}{$10^3\cdot R_{j,j+3}$} \\

\hline
1 & 0.15600 & 12.68 & 10.34 & 6.87 & 3.14 \\
2 & 0.15625 & 9.89 & 8.28 & 5.77 & 2.04 \\
3 & 0.15650 & 9.90 & 8.44 & 4.42 & 1.04 \\
4 & 0.15675 & 9.27 & 6.30 & 3.22 & 1.23 \\
5 & 0.15700 & 7.02 & 4.46 & 2.39 & \\
6 & 0.15725 & 4.34 & 2.82 &      & \\
7 & 0.15750 & 4.26 &      &      & \\

\hline

\end{tabular*}

\end{table*}



\begin{table*}[h]
\caption{\label{tab:error}Data and fit results for $P$ from 
standard runs and from tempering 
with 7 ensembles on the $12^4$ lattice.}

\vspace*{4mm}

\begin{tabular*}{16cm}{@{\extracolsep{\fill}}llll}

\hline
\multicolumn{1}{c}{$\kappa$} & 
\multicolumn{1}{c}{data} & 
\multicolumn{1}{c}{fit with full} &
\multicolumn{1}{c}{fit with diagonal} \\

\multicolumn{1}{c}{} &
\multicolumn{1}{c}{} &
\multicolumn{1}{c}{covariance matrix} &
\multicolumn{1}{c}{covariance matrix} \\

\hline
\hline
\multicolumn{4}{c}{standard HMC} \\
\hline
 0.15500 & 0.431825 (135) & & 0.432002 (112) \\
 0.15600 & 0.429872 (136) & & 0.429968 (66)  \\
 0.15650 & 0.429343 (132) & & 0.428952 (55)  \\
 0.15700 & 0.428309 (138) & & 0.427935 (59)  \\
 0.15750 & 0.426695 (91)  & & 0.426919 (77)  \\
 
\hline
\hline
\multicolumn{4}{c}{tempering (7 ensembles)} \\
\hline
 0.15600 & 0.430024 (79)  & 0.429973 (69)  & 0.430064 (50) \\
 0.15625 & 0.429494 (64)  & 0.429516 (56)  & 0.429578 (38) \\
 0.15650 & 0.429138 (63)  & 0.429060 (48)  & 0.429093 (30) \\
 0.15675 & 0.428749 (67)  & 0.428603 (48)  & 0.428607 (29) \\
 0.15700 & 0.428166 (81)  & 0.428146 (56)  & 0.428121 (35) \\
 0.15725 & 0.427560 (82)  & 0.427690 (68)  & 0.427636 (46) \\
 0.15750 & 0.427034 (102) & 0.427233 (84)  & 0.427150 (60) \\

\hline

\end{tabular*}

\end{table*}

\clearpage

\renewcommand{\textfraction}{0}

\clearpage

\newcommand{\rpic}[1]{\epsfig{file=rho.#1.eps,width=\WIDTH,%
bbllx=0,bblly=0,bburx=360,bbury=360}}
\newcommand{\Rpic}[1]{\epsfig{file=R.#1.eps,width=\WIDTH,%
bbllx=0,bblly=0,bburx=360,bbury=360}}

\renewcommand{\textfraction}{0}

\begin{figure}[h] 

\mbox{\rpic{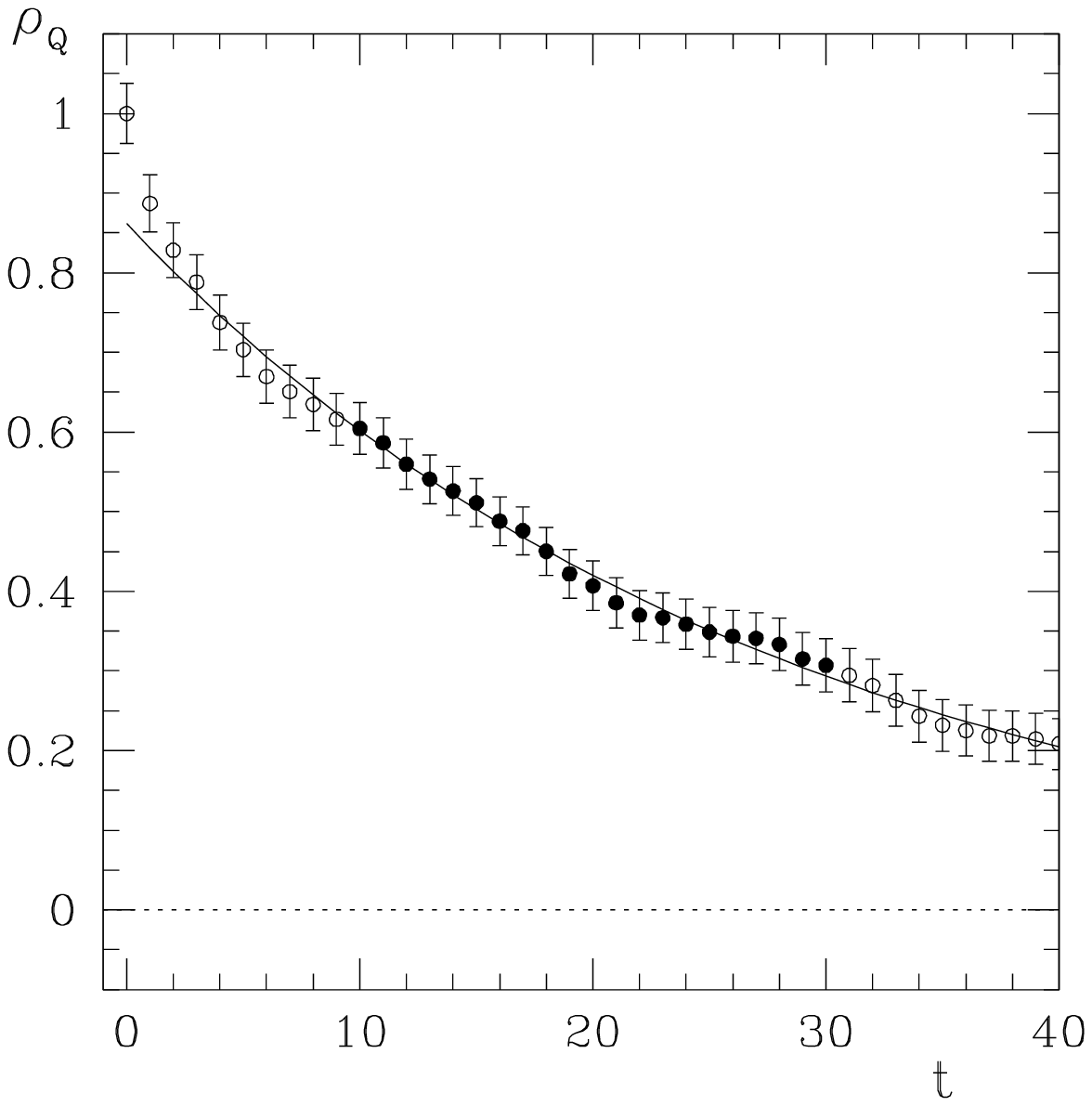}\hspace{\Width}\rpic{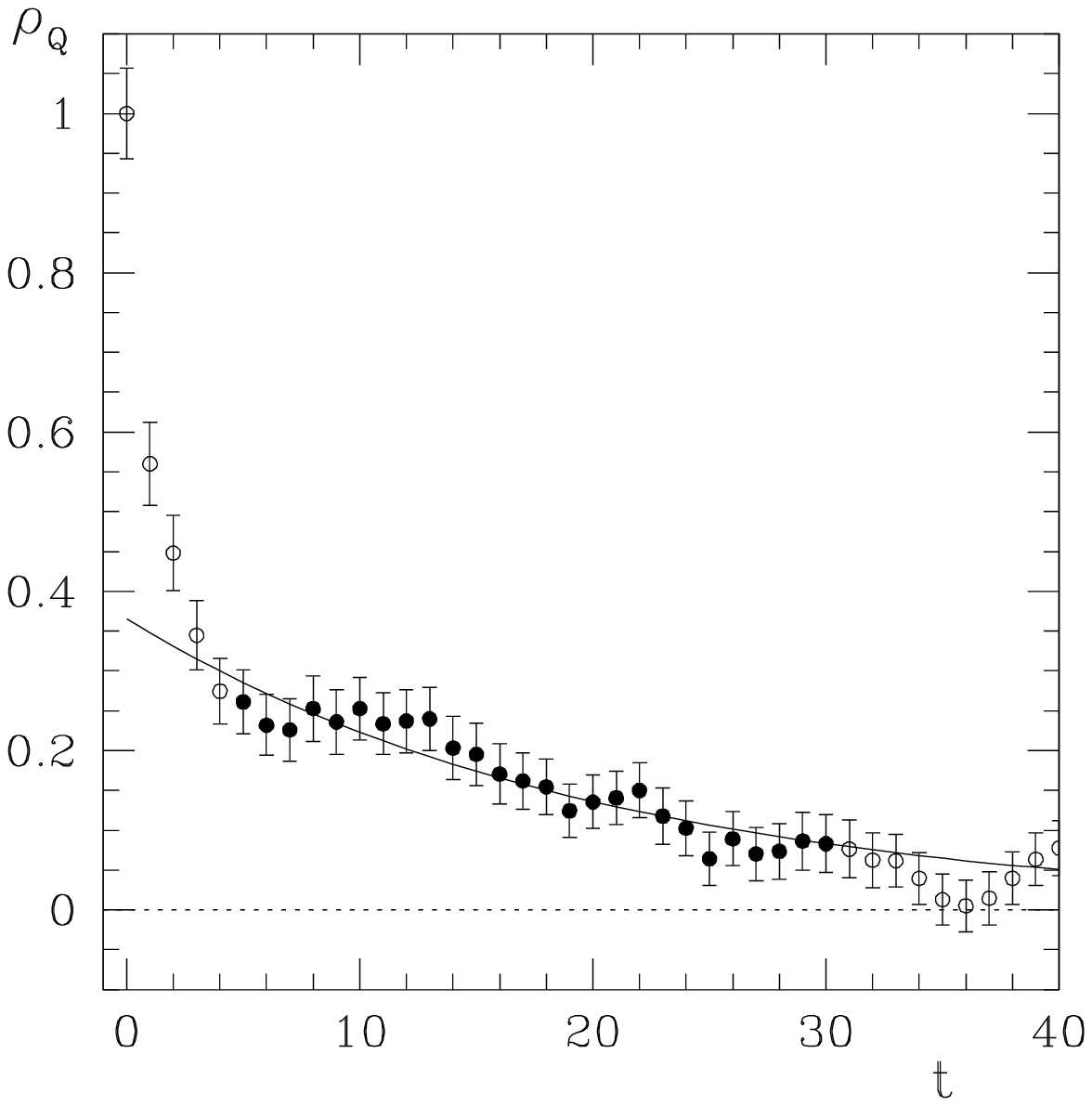}\hspace{\Width}\rpic{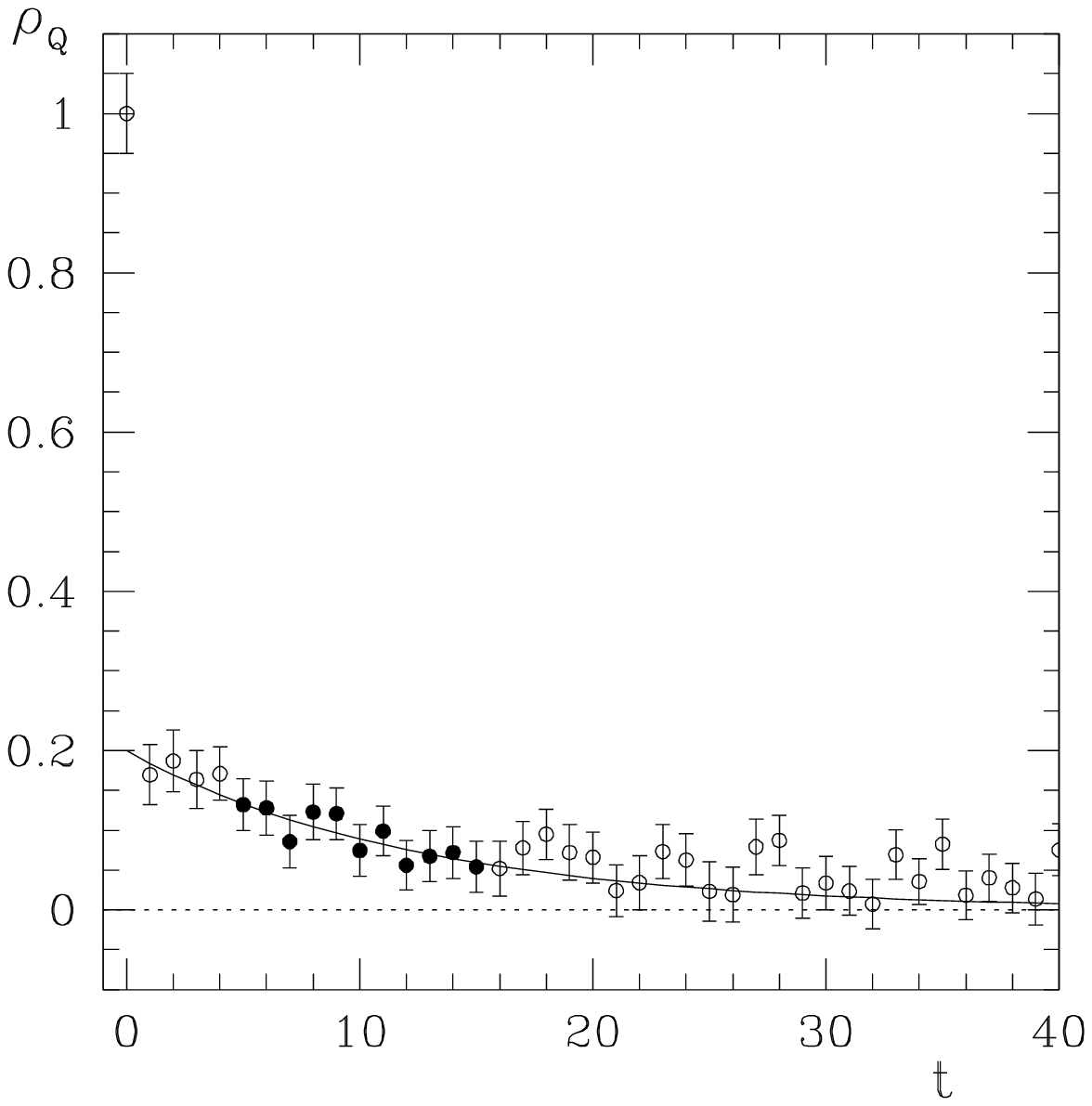}}

\smallskip

\caption{\label{fig:rho}Normalized autocorrelation functions for $Q$
for standard HMC (left) and tempering with 6 ensembles
(center) and 7 ensembles (right)
on a $12^4$ lattice for $\beta = 5.6$, $\kappa = 0.1565$. 
The errors indicated are the purely statistical ones.  The lines 
represent fits to the subset of data points with full symbols.}

\end{figure}

\renewcommand{\textfraction}{0}

\begin{figure}[h] 

\mbox{\Rpic{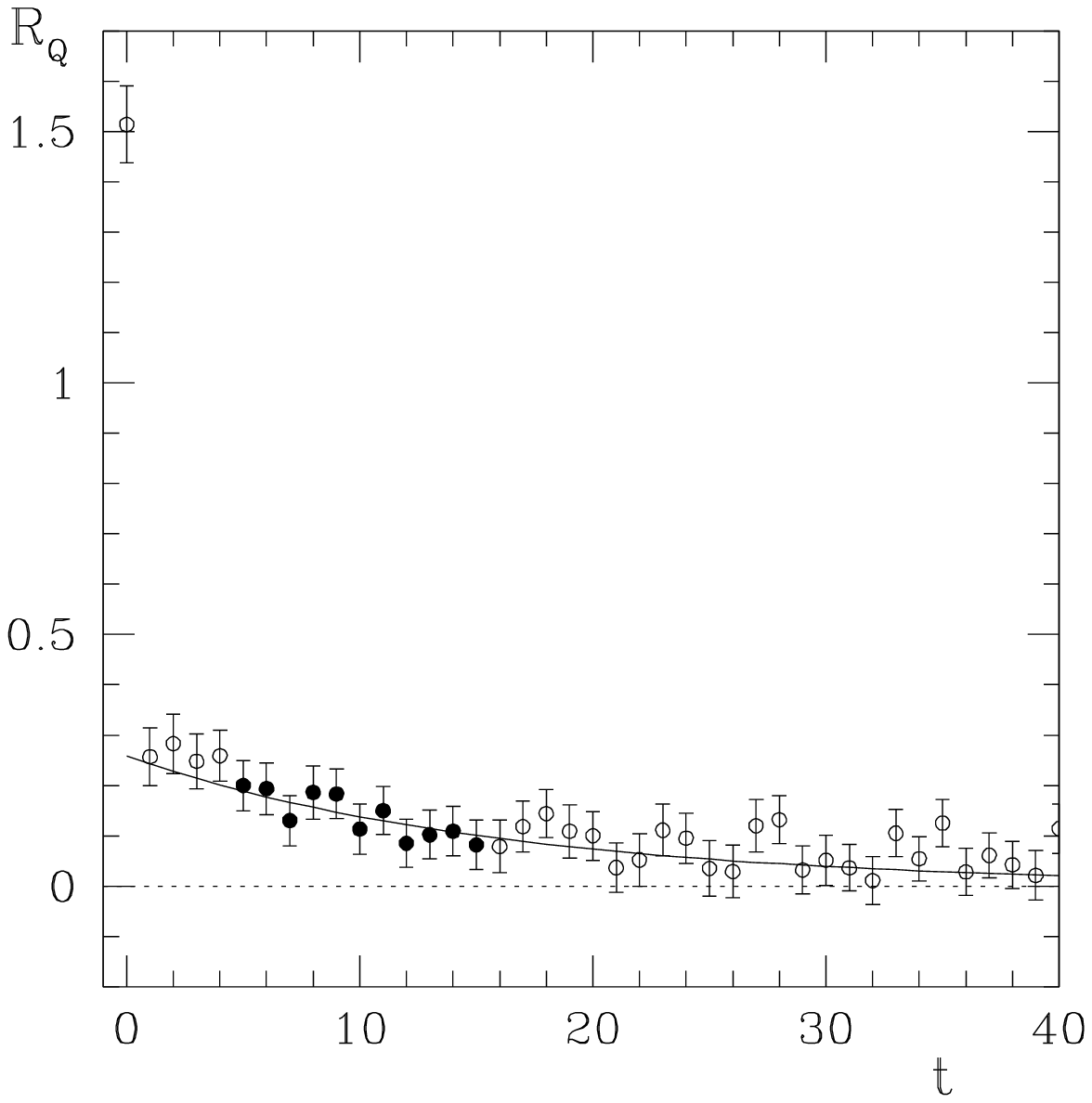}\hspace{\Width}\Rpic{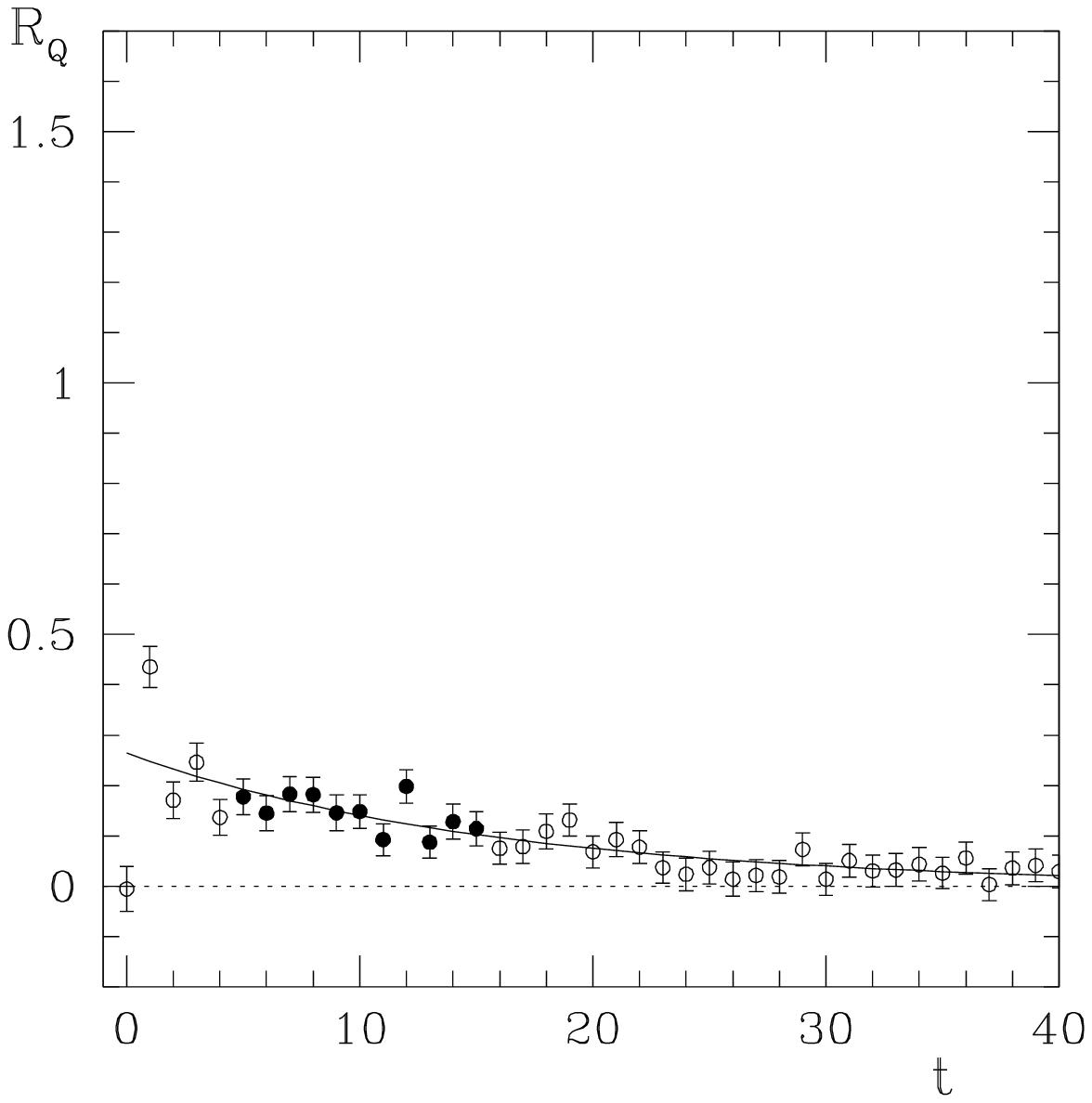}\hspace{\Width}\Rpic{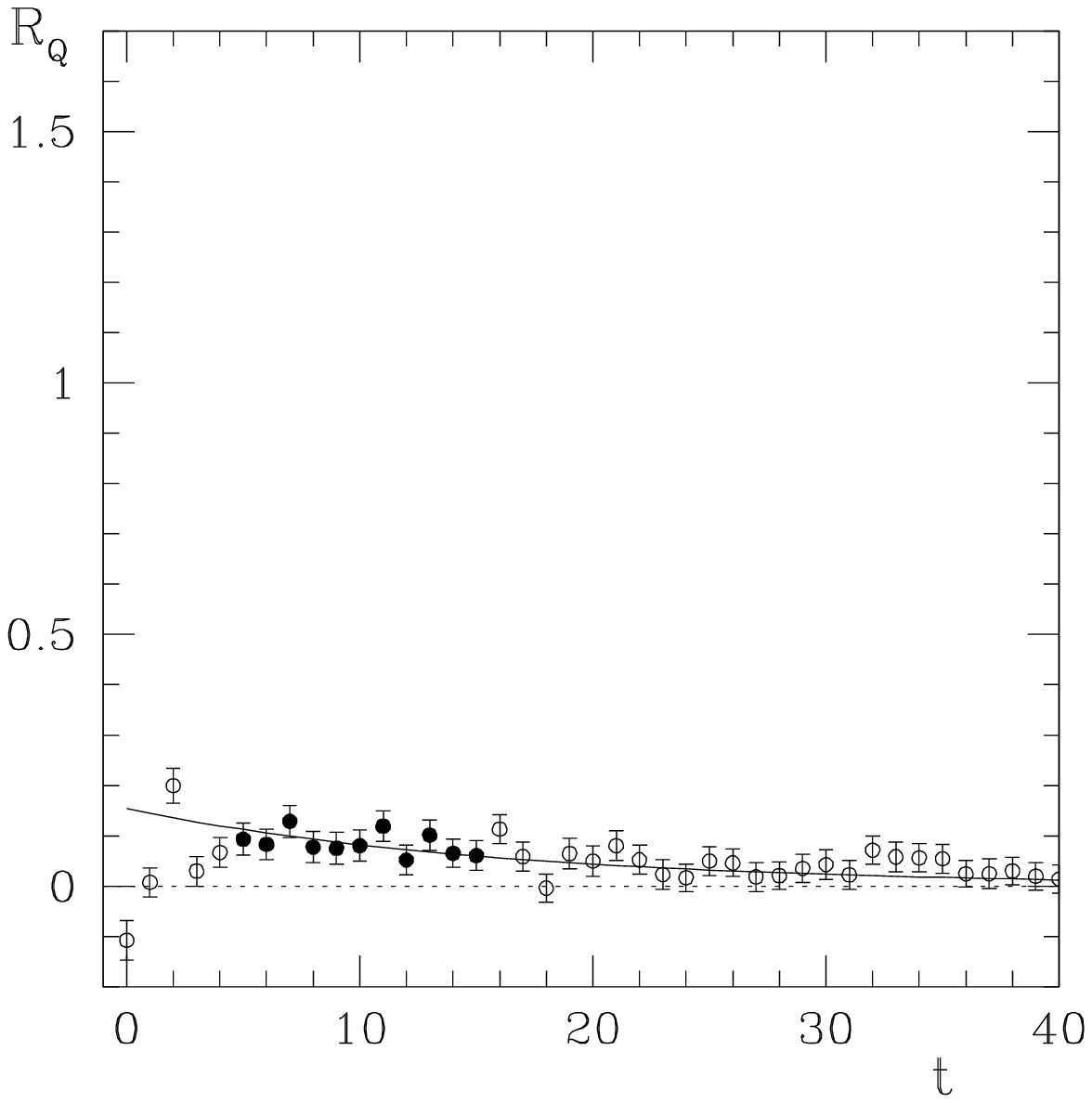}}

\smallskip

\caption{\label{fig:R}Correlation functions for $Q$ for tempering with 7 ensembles
on a $12^4$ lattice for $\beta = 5.6$. 
Shown are the autocorrelation function at $\kappa =
0.1565$ (left) and the cross-correlation functions for $\kappa =
0.1565, 0.15675$ (center) and $\kappa = 0.1565, 0.157$ (right).
The errors indicated are purely statistical ones.  The lines
represent one combined fit as explained in the text. Full data symbols
indicate the fit interval.}

\end{figure}

\end{document}